\documentclass[prd,%
preprint,%
tightenlines,%
amsfonts]{revtex4}
\usepackage{graphicx}

\begin{document}

\title{Tests of scalar-tensor gravity\footnote{Invited talk at the
workshop \textit{Phi in the Sky: The Quest for Cosmological Scalar
Fields}, Porto, July 2004}}

\date{September 20, 2004}

\author{Gilles \surname{Esposito-Far\`ese}}
\affiliation{${\mathcal{G}}{\mathbb{R}}
\varepsilon{\mathbb{C}}{\mathcal{O}}$, FRE 2435-CNRS, Institut
d'Astrophysique de Paris, 98bis boulevard Arago, F-75014
Paris, France}

\begin{abstract}
The best motivated alternatives to general relativity are
scalar-tensor theories, in which the gravitational interaction is
mediated by one or several scalar fields together with the usual
graviton. The analysis of their various experimental constraints
allows us to understand better which features of the models have
actually been tested, and to suggest new observations able to
discriminate between them. This talk reviews three classes of
constraints on such theories, which are qualitatively different from
each other: (i)~solar-system experiments; (ii)~binary-pulsar tests
and future detections of gravitational waves from inspiralling
binaries; (iii)~cosmological observations. While classes (i) and
(ii) impose precise bounds respectively on the first and second
derivatives of the matter-scalar coupling function, (iii) \textit{a
priori} allows us to reconstruct the full shapes of the functions of
the scalar field defining the theory, but obviously with more
uncertainties and/or more theoretical hypotheses needed. Simple
arguments such as the absence of ghosts (to guarantee the stability
of the field theory) nevertheless suffice to rule out a wide class
of scalar-tensor models. Some of them can be probed only if one
takes simultaneously into account solar-system and cosmological
observations.
\end{abstract}

\preprint{gr-qc/0409081}

\maketitle

\section{Introduction}
\label{introduction}
In the most natural alternative theories to general relativity (GR),
gravity is mediated not only by a (spin-2) graviton corresponding to
a metric $g_{\mu\nu}$, but also by a (spin-0) scalar field
$\varphi$. Such scalar partners generically arise in all
extra-dimensional theories, and notably in string theory. A dilaton
is indeed already present in the supermultiplet of the
10-dimensional graviton, and several other scalar fields (called the
moduli) also appear when performing a Kaluza-Klein dimensional
reduction to our usual spacetime. They correspond to the components
of the metric tensor $g_{mn}$ in which $m$ and $n$ label extra
dimensions. Moreover, contrary to other alternative theories of
gravity, scalar-tensor theories respect most of GR's symmetries:
conservation laws, constancy of non-gravitational constants, and
local Lorentz invariance even if a subsystem is influenced by
external masses. They can also satisfy exactly the weak equivalence
principle (universality of free fall of laboratory-size objects)
even for a massless scalar field.

Scalar fields are also involved in the cosmological models which
reproduce most consistently present observational data. In
particular, inflation theory is based on the presence of a scalar
$\varphi$ in a potential $V(\varphi)$ (for instance parabolic). It
behaves as a fluid with a positive energy density $8\pi G
\rho_\varphi = \dot\varphi^2 + 2 V(\varphi)$ but a negative pressure
$8\pi G p_\varphi = \dot\varphi^2 - 2 V(\varphi)$. This causes a
period of exponential expansion of the universe, which can explain
why causally disconnected regions at present may have been connected
long ago. The isotropy of the observed cosmic microwave background
(CMB) can thus be understood. Inflation also predicts that our
universe is almost spatially flat, because any initial curvature has
been exponentially reduced by the expansion. This is in remarkable
agreement with the location of the first acoustic peak of the CMB
spectrum at a multipolar index $\ell \simeq 220$~\cite{wmap03}.
Various observations, notably of type Ia supernovae~\cite{SNIa},
tell us that there is about 70\% of negative-pressure dark energy in
our present universe ($\Omega_\Lambda \simeq 0.7$), suggesting that
its expansion has been re-accelerating recently, since redshifts $z
\sim 1$. This can be explained by the presence of a cosmological
constant $\Lambda$ in GR, but the quantity $\Omega_\Lambda \simeq
0.7$ translated in natural units gives an extremely small value
$\Lambda \simeq 3\times 10^{-122} c^3/(\hbar G)$, very problematic
for particle physics if $\Lambda$ is to be interpreted as the vacuum
energy. This is the main reason why ``quintessence'' models have
been proposed, in which the cosmological constant is replaced again
by the potential $V(\varphi)$ of a scalar field. Its evolution
towards a minimum of $V$ during the cosmological expansion then
explains more naturally why the present value $V(\varphi_0) \simeq
\Lambda/2$ is so small.

Besides these theoretical and experimental reasons for studying
scalar-tensor theories of gravity, one of their greatest interests
is to embed GR within a class of mathematically consistent
alternatives, in order to understand better which theoretical
features have been experimentally tested, and which can be tested
further. To simplify this review, we will restrict it to models
involving a single scalar field, although the study of
tensor--multi-scalar theories can also be done in great
detail~\cite{def1}. It suffices to note that their phenomenology is
richer but similar to the single scalar case, at least when all
scalar degrees of freedom carry positive energy, as required for the
vacuum to be stable. In view of the very precise experimental
verifications of the weak equivalence principle, we will also focus
this review on theories which satisfy it exactly. In other words, we
will assume that the action of geometry on matter is the same as in
GR, but that the dynamics of geometry and the action of matter on it
is modified because of the presence of a scalar field (see
J.P.~Uzan's contribution to the present proceedings, and in
particular the upper-right panel of its Fig.~1). All matter fields,
including gauge bosons, will thus be assumed to be universally
coupled to one second rank symmetric tensor, say $\widetilde
g_{\mu\nu}$. The difference with GR lies in the kinetic term of this
tensor, which is now a mixing of spin-2 and spin-0 excitations. One
indeed defines $\widetilde g_{\mu\nu} \equiv A^2(\varphi)
g_{\mu\nu}$, where $A(\varphi)$ is a function of the scalar field,
while $g_{\mu\nu}$ denotes the Einstein (spin-2) metric. We will
mainly consider the following action~\cite{BNW},
\begin{equation}
S = \frac{c^3}{4 \pi G}
\int\sqrt{-g}\left\{\frac{R}{4}
-\frac{1}{2}(\partial_\mu\varphi)^2
-V(\varphi)\right\}
+ S_{\rm matter}\left[{\rm matter} ; \widetilde g_{\mu\nu} \equiv
A^2(\varphi) g_{\mu\nu}\right],
\label{action}
\end{equation}
which depends on two functions of the scalar field: the coupling
function $A(\varphi)$ to matter, and a potential $V(\varphi)$. [Our
signature is $\scriptstyle -+++$, $R$ is the scalar curvature of
$g_{\mu\nu}$, and $g$ its determinant.]

In Section~\ref{localTests}, we will review the main experimental
constraints on this coupling function $A(\varphi)$, coming from two
qualitatively different classes of data: solar-system experiments
and binary-pulsar observations~\cite{def7and9,MGX}. In
Section~\ref{reconstruction}, we will underline that cosmological
observations give again very different types of constraints:
Although they are more noisy, they \textit{a priori} allow the
reconstruction of both $A(\varphi)$ and
$V(\varphi)$~\cite{beps00,efp01}. The most general scalar-tensor
theories will be defined in Sec.~\ref{general}, and we will
illustrate in Sec.~\ref{GaussBonnet}, on a very special but
interesting case, that one must sometimes take simultaneously into
account solar-system and cosmological observations to obtain
significant constraints on a theory~\cite{moriond03}. We will
finally give our conclusions in Sec.~\ref{conclusions}.

\section{Solar-system and binary-pulsar constraints}
\label{localTests}
\subsection{Solar-system tests}
A massive scalar field has a negligible effect on the motion of
celestial bodies if its mass is large with respect to the inverse of
the interbody distances. On the other hand, if its mass is small
enough, its potential $V(\varphi)$ can be locally neglected, but its
coupling function to matter, $A(\varphi)$, is strongly constrained
by experiment, as we will see below. The intermediate case, for
which the size of the experiment is of the same order as the inverse
scalar mass, is the main topic of E.~Adelberger's contribution to
the present proceedings (see also \cite{Adelberger}). The
``chameleon'' field~\cite{Khoury} is another case where both the
potential and the matter--scalar coupling function must be taken
into account in solar-system experiments; see P.~Brax's contribution
to these proceedings. In the present section, we will review the
constraints on $A(\varphi)$ when the potential $V(\varphi)$ can be
neglected.

The precision of solar-system observations allow us not only to test
Newton's law, but also its relativistic corrections $\propto 1/c^2$,
called ``post-Newtonian''. At this order, the predictions of metric
theories of gravity can be parametrized by a set of 10 real numbers
in the so called ``PPN'' formalism (parametrized
post-Newtonian)~\cite{PPN}. All of them are presently constrained to
be very close to their general relativistic values~\cite{w01}, and
in particular the two famous parameters $\beta^{\rm PPN}$ and
$\gamma^{\rm PPN}$ introduced by Eddington in the Schwarzschild
metric:
\begin{eqnarray}
-g_{00}&=& 1 - 2\frac{Gm}{rc^2} + 2 \beta^{\rm PPN}
\left(\frac{Gm}{rc^2}\right)^2 +
\mathcal{O}\left(\frac{1}{c^6}\right)\,,\nonumber\\
g_{ij}&=&\delta_{ij}\left(1+2\gamma^{\rm PPN}\frac{Gm}{rc^2}\right)
+ \mathcal{O}\left(\frac{1}{c^4}\right)\,.
\label{Edd}
\end{eqnarray}
General relativity corresponds to $\beta^{\rm PPN} = \gamma^{\rm
PPN} = 1$. In scalar-tensor theories~\cite{def1,def7and9}, these two
latter parameters are the only ones which can differ from their GR
values. They are related to the first two derivatives of
\begin{equation}
\ln A(\varphi) \equiv
\alpha_0(\varphi-\varphi_0)
+\frac{1}{2}\beta_0 (\varphi-\varphi_0)^2
+ \mathcal{O}(\varphi-\varphi_0)^3,
\label{lnA}
\end{equation}
computed at the background value $\varphi_0$ of the scalar field:
\begin{equation}
\gamma^{\rm PPN}-1 = - \frac{2 \alpha_0^2}{1+\alpha_0^2}\,,
\qquad \beta^{\rm PPN}-1 = \frac{1}{2}\,
\frac{\alpha_0\beta_0\alpha_0}{(1+\alpha_0^2)^2}\,.
\label{PPN}
\end{equation}
\begin{figure}
\includegraphics[height=.4\textheight]{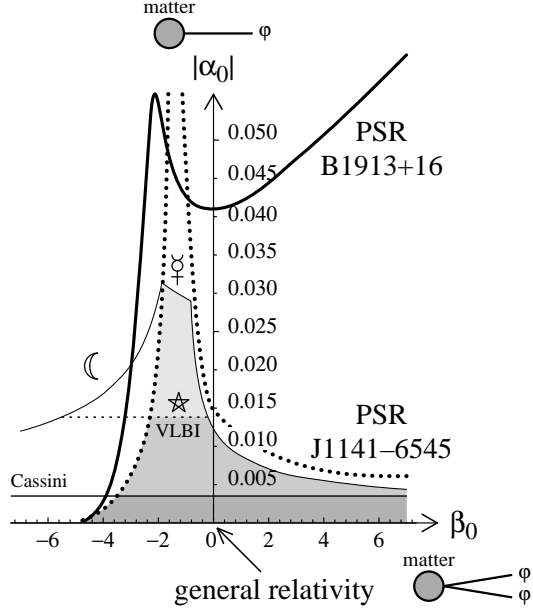}
\caption{Solar-system and binary-pulsar constraints on the
matter-scalar coupling function $\ln A(\varphi) =
\alpha_0(\varphi-\varphi_0) + \frac{1}{2} \beta_0
(\varphi-\varphi_0)^2 + \mathcal{O}(\varphi-\varphi_0)^3$. The
allowed region is shaded. The vertical axis ($\beta_0=0$)
corresponds to Brans-Dicke theory with a parameter $2\omega_{\rm
BD}+3 = 1/\alpha_0^2$. The horizontal axis ($\alpha_0=0$)
corresponds to theories which are perturbatively equivalent to GR,
i.e., which predict strictly no deviation from it (at any order
$1/c^n$) in the weak-field conditions of the solar system.
\label{fig1}}
\end{figure}
The factor $\alpha_0^2$ comes from the exchange of a scalar particle
between two bodies, whereas $\alpha_0\beta_0\alpha_0$ comes from a
scalar exchange between three bodies. Solar system experiments
impose the constraints displayed as thin lines in Fig.~\ref{fig1}.
The Mercury symbol refers to the perihelion shift of this planet,
whose observed value imply the bound~\cite{Mercury}
\begin{equation}
\vert 2 \, \gamma^{\rm PPN} - \beta^{\rm PPN} -1 \vert
< 3\times10^{-3},
\label{Mercury}
\end{equation}
the Moon symbol refers to Lunar Laser Ranging~\cite{LLR}:
\begin{equation}
4 \, \beta^{\rm PPN} -\gamma^{\rm PPN} -3
= (-0.7\pm 1) \times 10^{-3},
\label{LLR}
\end{equation}
the star symbol to light deflection as measured by Very Long
Baseline Interferometry~\cite{VLBI}:
\begin{equation}
\vert \gamma^{\rm PPN} -1 \vert < 4\times10^{-4},
\end{equation}
and the label ``Cassini'' to the impressive recent constraint
obtained by measuring the time delay variation to the Cassini
spacecraft near solar conjunction~\cite{Cassini}:
\begin{equation}
\gamma^{\rm PPN} -1 = (2.1\pm 2.3) \times 10^{-5}.
\label{Cassini}
\end{equation}
Solar-system tests thus constrain the first derivative $\alpha_0$ to
be small, i.e., the linear interaction between matter and the scalar
field to be weak. On the other hand, the second derivative $\beta_0$
may take large values (of either sign), i.e., matter may be strongly
coupled to two scalar lines.

\subsection{Nonperturbative strong-field effects}
\label{nonpert}
\begin{figure}
\includegraphics[height=.325\textheight]{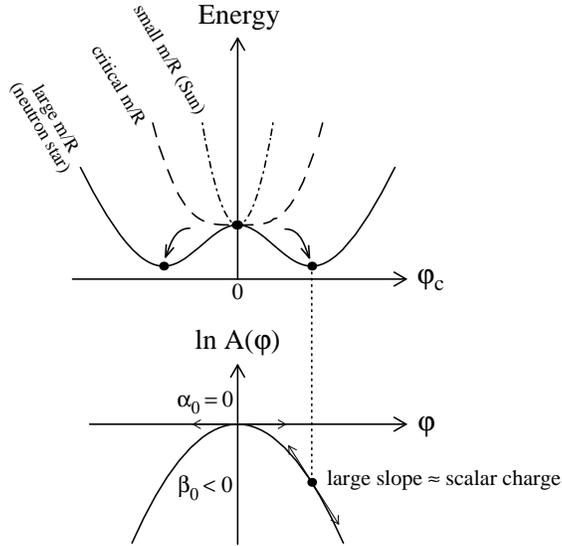}
\caption{Heuristic argument to explain the phenomenon of
``spontaneous scalarization''. When $\beta_0 <0$ and the compactness
$Gm/Rc^2$ of a body is large enough, it is energetically favorable
to create a local scalar field different from the background value.
The body becomes thus strongly coupled to the scalar field.
\label{fig2}}
\end{figure}
At higher post-Newtonian orders $1/c^n$, a simple diagrammatic
argument shows that any deviation from GR involves at least two
factors $\alpha_0$, and has the schematic form
\begin{equation}
\hbox{deviation from GR} = \alpha_0^2\times
\left[\lambda_0 + \lambda_1 \frac{Gm}{Rc^2} + \lambda_2
\left(\frac{Gm}{Rc^2}\right)^2+\cdots \right],
\label{8}
\end{equation}
where $m$ and $R$ denote the mass and radius of the considered body,
and $\lambda_0$, $\lambda_1$, \dots\ are constants built from the
coefficients $\alpha_0$, $\beta_0$, \dots\ of expansion~(\ref{lnA}).
Since $\alpha_0^2$ is experimentally known to be small, we thus
expect the theory to be close to GR at any order. However, some
nonperturbative effects may occur in strong-field conditions: If the
compactness $Gm/Rc^2$ of a body is greater than a critical value,
the square brackets of Eq.~(\ref{8}) can become large enough to
compensate even a vanishingly small $\alpha_0^2$. This can happen
notably for neutron stars, whose compactnesses are of order $Gm/Rc^2
\sim 0.2$, as compared to only $2\times 10^{-6}$ for the Sun or
even $7\times 10^{-10}$ for the Earth. To illustrate this, let us
consider a model for which $\alpha_0$ vanishes strictly, i.e., which
is perturbatively equivalent to GR: There is strictly no deviation
{}from GR at any order in a perturbative expansion in powers of
$1/c$. A parabolic coupling function $\ln A(\varphi) = \frac{1}{2}
\beta_0 \varphi^2$ suffices for our purpose (we set here $\varphi_0
= 0$ to simplify). At the center of a static body, the scalar field
takes a particular value $\varphi_c$, and it decreases as $1/r$
outside. The energy of such a scalar field configuration involves
two contributions, coming respectively from the kinetic term and
{}from the matter-scalar coupling function in action~(\ref{action}).
As a rough estimate of its value, one can write
\begin{equation}
{\rm Energy} \simeq \int\left[\frac{1}{2}(\partial_i\varphi)^2
+\rho\, e^{\beta_0\varphi^2/2} \right]
\simeq mc^2\left(
\frac{\varphi_c^2/2}{Gm/Rc^2}
+ e^{\beta_0\varphi_c^2/2}\right).
\label{9}
\end{equation}
When $\beta_0<0$, this is the sum of a parabola and a Gaussian, and
if the compactness $Gm/Rc^2$ is large enough, the function ${\rm
Energy}(\varphi_c)$ has the shape of a Mexican hat, see
Fig.~\ref{fig2}. The value $\varphi_c =0$ now corresponds to a local
\textit{maximum} of the energy. It is therefore energetically
favorable for the star to create a nonvanishing scalar field
$\varphi_c$, and thereby a nonvanishing ``scalar charge'' $d \ln
A(\varphi_c)/d\varphi_c = \beta_0\varphi_c$. This phenomenon is
analogous to the spontaneous magnetization of ferromagnets.

\begin{figure}
\includegraphics[height=.225\textheight]{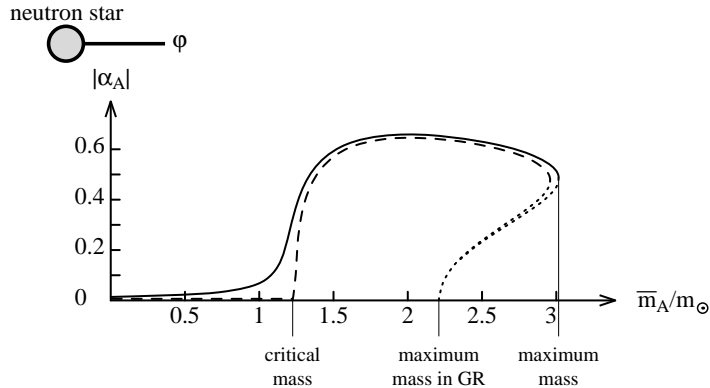}
\caption{Scalar charge $\alpha_{\rm A}$ versus baryonic mass
$\overline m_{\rm A}$, for the model $A(\varphi) = \exp(-3\varphi^2)$
(i.e., $\beta_0 = -6$). The solid line corresponds to the maximum
value of $\alpha_0$ allowed by Eqs.~(\ref{Mercury})-(\ref{LLR}), and
the dashed line to $\alpha_0 = 0$. The dotted lines correspond to
unstable configurations of the star.
\label{fig3}}
\end{figure}
This heuristic argument has been verified by explicit numerical
calculations, taking into account the coupled differential equations
of the metric and the scalar field, and using various realistic
equations of state to describe nuclear matter inside a neutron
star~\cite{def7and9}. The correct definition of the linear coupling
strength between a compact body ${\rm A}$ and the scalar field reads
$\alpha_{\rm A}\equiv \partial\ln m_{\rm A}/\partial\varphi_0$. It
is plotted in Fig.~\ref{fig3} for the particular model $\beta_0 =
-6$. One finds that there exists indeed a ``spontaneous
scalarization'' above a critical mass (whose value decreases as
$-\beta_0$ grows).

On the other hand, if $\beta_0 > 0$, both the above heuristic
argument and the actual numerical calculations show that
$|\alpha_{\rm A}| < |\alpha_0|$. In that case, one finds that
neutron stars are even less coupled to the scalar field than
solar-system bodies. This decoupling is similar to the behavior of
the ``chameleon'' field~\cite{Khoury}, but it is here caused by the
nonlinearity of $\ln A(\varphi)$, whereas the chameleon's decoupling
is due to a competition between $A(\varphi)$ and the potential
$V(\varphi)$.

\subsection{Binary-pulsar tests}
The scalar charge $\alpha_{\rm A}$ enters the predictions of the
theory in the same way as $\alpha_0$ in weak-field conditions. For
instance, in the orbital motion of two bodies ${\rm A}$ and ${\rm
B}$, the Eddington parameter $\gamma^{\rm PPN}$ keeps the form of
Eq.~(\ref{PPN}), but it now involves the product $\alpha_{\rm
A}\alpha_{\rm B}$ of the two scalar charges instead of $\alpha_0^2$.
Similarly, the strong-field analogue of $\beta^{\rm PPN}$ involves
products of scalar charges and their derivatives $\beta_{\rm
A}\equiv\partial\alpha_{\rm A}/\partial\varphi_0$. Since
$|\alpha_{\rm A}| \simeq 0.6$ in the model of Fig.~\ref{fig3}, one
thus expects deviations by $\sim 35\%$ from some general relativistic
predictions. Moreover, the quadratic coupling strength $\beta_{\rm A}$
can take very large numerical values near the critical mass, like the
magnetic susceptibility of ferromagnets. Therefore, even larger
deviations from GR are found when the mass of a neutron star happens
to be close to the critical one.

\begin{figure}
\includegraphics[height=.29\textheight]{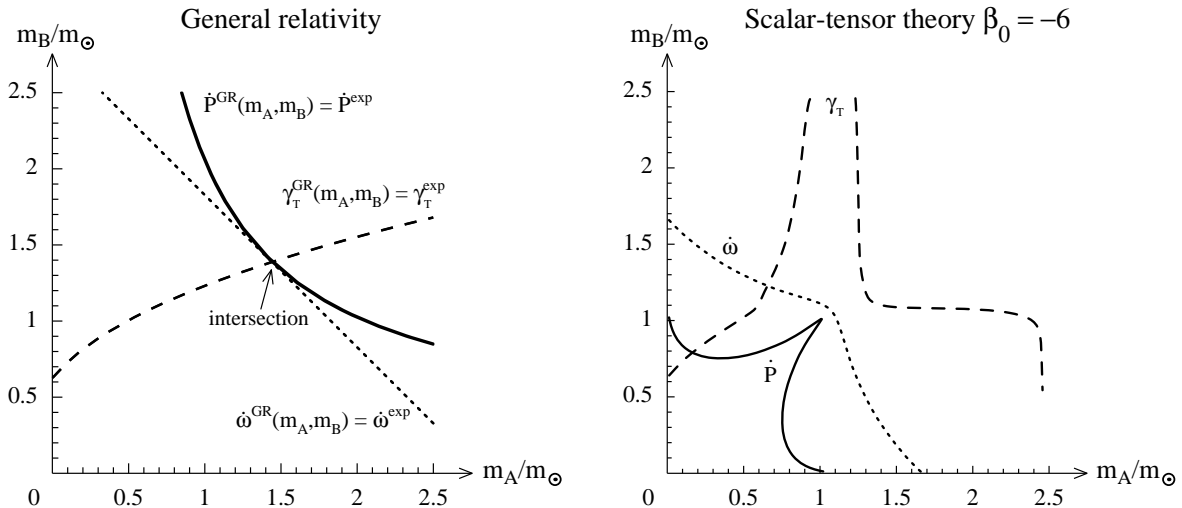}
\caption{Mass plane ($m_{\rm A}=$ pulsar, $m_{\rm B} =$ companion)
of the Hulse-Taylor binary pulsar PSR B1913+16 in general relativity
(left panel) and for a scalar-tensor theory with $\beta_0 = -6$
(right panel). The widths of the lines are larger than $1\sigma$
error bars. While GR passes the test with flying colors, the value
$\beta_0 = -6$ is ruled out.
\label{fig4}}
\end{figure}
Binary pulsars are ideal tools for testing gravity in strong-field
conditions. A pulsar is a rapidly rotating neutron star emitting a
beam of radio waves, like a lighthouse. Experiment tells us that
isolated pulsars are very stable clocks, when they are old enough. A
pulsar ${\rm A}$ orbiting a companion ${\rm B}$ is thus a moving
clock, the best tool that one could dream of to test a relativistic
theory. Indeed, by precisely timing its pulse arrivals, one gets a
stroboscopic information on its orbit, and one can measure several
relativistic effects. Such effects do depend on the two masses
$m_{\rm A}$, $m_{\rm B}$, which are not directly measurable.
However, two different effects suffice to determine them, and a
third relativistic observable then gives a test of the theory. In
the case of the famous Hulse-Taylor binary pulsar PSR B1913+16,
three relativistic parameters have been determined with great
accuracy~\cite{1913}: (i)~the Einstein time delay parameter
$\gamma_T$, which combines the second-order Doppler effect ($\propto
v_{\rm A}^2/2 c^2$, where $v_{\rm A}$ is the pulsar's velocity)
together with the redshift due to the companion ($\propto G m_{\rm
B}/r_{\rm AB} c^2$, where $r_{\rm AB}$ the pulsar-companion
distance); (ii)~the periastron advance $\dot\omega$ (relativistic
effect of order $v^2/c^2$); and (iii)~the rate of change of the
orbital period, $\dot P$, caused by gravitational radiation damping
(an effect of order $v^5/c^5$ in GR, but of order $v^3/c^3$ in
scalar-tensor theories; see below). Figure~\ref{fig4} displays the
plane of the two \textit{a priori} unknown masses $m_{\rm A}$ and
$m_{\rm B}$. For each relativistic parameter, the prediction of a
given theory is consistent with the observed value only along a thin
line. In GR, the fact that the three lines meet in one point means
that there exists a pair of masses ($m_{\rm A}, m_{\rm B}$)
simultaneously consistent with all three observables. This is thus a
spectacular confirmation of Einstein's theory.

Obviously, these lines are deformed in scalar-tensor theories, and
if they do not have any common intersection, the corresponding model
is ruled out. The right panel of Fig.~\ref{fig4} illustrates such a
case. The allowed theories lie below and to the right of the line
labeled PSR B1913+16 in Fig.~\ref{fig1}. This plot shows vividly the
qualitative difference between solar-system and binary-pulsar
observations. Indeed, the latter impose
\begin{equation}
\beta_0 > -4.5\,,
\label{beta0min}
\end{equation}
even for a vanishingly small $\alpha_0$. This constraint is due to
the spontaneous scalarization of neutron stars, which occurs when
$-\beta_0$ is large enough. Equations~(\ref{PPN}) allow us to
rewrite this inequality in terms of the Eddington parameters
$\beta^{\rm PPN}$ and $\gamma^{\rm PPN}$, which are both consistent
with 1 in the solar system. One finds
\begin{equation}
\frac{\beta^{\rm PPN}-1}{\gamma^{\rm PPN}-1} < 1.1\,.
\end{equation}
The singular ($0/0$) nature of this ratio underlines why such a
conclusion could not be obtained in weak-field experiments.

Several other binary pulsars are presently known and accurately
timed. To simplify Fig.~\ref{fig1}, we did not display all the
constraints they impose in the theory plane (see Ref.~\cite{MGX}).
We just plotted as a dotted line the tight ones imposed by the
recently timed neutron star--white dwarf binary PSR
J1141$-$6545~\cite{1141}. Contrary to neutron stars, the white dwarf
companion is never spontaneously scalarized, because its compactness
$Gm/Rc^2$ is too small. This binary system is thus very
asymmetrical, and the pulsar's scalar charge $\alpha_{\rm A}$ is
generically much greater than its companion's, $\alpha_{\rm B}$.
This causes a huge deviation from GR in the orbital period variation
$\dot P$. Indeed, the energy flux carried out by gravitational waves
is of the form
\begin{eqnarray}
\hbox{Energy flux} = \left\{\frac{{\rm Quadrupole}}{c^5} +
\mathcal{O}\left(\frac{1}{c^7}\right)\right\}_{\rm helicity~2}
\hskip 5.7cm
\nonumber\\
+ \left\{\frac{{\rm Monopole}}{c}\left(0+\frac{1}{c^2}\right)^2
+ \frac{{\rm Dipole}}{c^3} (\alpha_{\rm A}-\alpha_{\rm B})^2
+ \frac{{\rm Quadrupole}}{c^5}
+ \mathcal{O}\left(\frac{1}{c^7}\right)\right\}_{\rm
helicity~0}
\label{11}
\end{eqnarray}
The first curly brackets contain the prediction of general
relativity, of order $v^5/c^5$, whereas the second ones contain the
extra contributions predicted in tensor-scalar theories. In
particular, the dipolar contribution is of order $v^3/c^3$, much
larger than the usual quadrupole of GR if the two scalar charge are
significantly different. This is the reason why the asymmetrical
system PSR J1141$-$6545 is so constraining for scalar-tensor
theories, although its experimental uncertainties on $\dot P$ are
still rather large. Notice that this binary pulsar is almost as
constraining as solar-system tests even in the region $\beta_0 > 0$;
see Fig.~\ref{fig1}. It should probe values of the Eddington
parameter $|\gamma^{\rm PPN}-1| \sim 10^{-6}$ by the end of the
decade, i.e., an order of magnitude better than present solar-system
limits~(\ref{Cassini}).

The LIGO and VIRGO interferometers will detect gravitational waves
emitted by inspiralling binaries. Since matter is known to be weakly
coupled to the scalar field in the solar system (small value of
$|\alpha_0|$), these detectors will not be sensitive of the
helicity-0 waves. On the other hand, the time evolution of the
helicity-2 chirp does depend on the energy flux~(\ref{11}), which
differs significantly from the GR prediction when the scalar charges
$\alpha_{\rm A}$, $\alpha_{\rm B}$ are nonzero. Therefore, the GR
wave templates used for matched filtering in LIGO and VIRGO may not
be accurate if there exists a scalar partner to the graviton, and
the signal-to-noise ratio may then drop. Fortunately, it was shown
in \cite{def7and9} that binary-pulsar data are so precise that they
already exclude the models which would have predicted significant
effects in the gravitational waveforms. Therefore, although these
interferometers are \textit{a priori} more sensitive to the scalar
field than classic solar-system tests, one may securely use the GR
wave templates for their data analysis. On the other hand, it was
shown in~\cite{will02} that the LISA space interferometer can be
sensitive to scalar effects which are still allowed by all present
tests. Future binary-pulsar data should nevertheless probe them
before the LISA mission is launched~\cite{MGX}.

In conclusion, solar-system tests tightly constrain the first
derivative of $\ln A(\varphi)$ (linear matter-scalar coupling
strength $\alpha_0$), whereas binary-pulsar data impose that its
second derivative $\beta_0$ (quadratic coupling
matter-scalar-scalar) is not large and negative. We will now see
that cosmological observations give access to the full shape of this
coupling function, of course not with the same accuracy as the above
tests, but with the capability of constraining any higher derivative
of $\ln A(\varphi)$ (vertex of matter with any number of scalar
lines). Moreover, cosmological data can also give access to the full
shape of the potential $V(\varphi)$.

\section{Reconstruction of a scalar-tensor theory from
cosmological observations}
\label{reconstruction}
In cosmology, the usual approach to study quintessence models
is to assume a particular form for the potential $V(\varphi)$
(and the matter-scalar coupling function $A(\varphi)$ when
one considers ``extended quintessence'' models), to compute
all possible observable predictions, and to compare them to
experimental data.

In contrast, in the phenomenological approach, one wishes to
\textit{reconstruct} the Lagrangian of the theory from cosmological
observations. It was proved in~\cite{beps00} that the knowledge of
the luminosity distance $D_L(z)$ and of the density fluctuations
$\delta_m(z) = \delta\rho/\rho$ as functions of the redshift $z$
indeed suffices to reconstruct both the potential $V(\varphi)$ and
the coupling function $A(\varphi)$. [The knowledge of the Hubble
constant $H_0$ or the present matter density $\Omega_{m,0}$ is not
even necessary, as they derive from $D_L(z)$ and $\delta_m(z)$.] In
order to write the corresponding equations as simply as possible, it
is convenient to perform a change of variables on
action~(\ref{action}). Let us define
\begin{eqnarray}
{\widetilde g_{\mu\nu}} \equiv
A^2(\varphi)g_{\mu\nu}\,;&&{\Phi}
\equiv A^{-2}(\varphi)\,;
\nonumber\\
2{\omega_{\rm BD}(\Phi)}
+ 3 \equiv A^2(\varphi)/A'^2(\varphi)\,;&&{U(\Phi)}
\equiv 2V(\varphi)/A^4(\varphi)\,.
\label{BransDickeVars}
\end{eqnarray}
Then action~(\ref{action}) takes the ``Brans-Dicke'' form
\begin{eqnarray}
S = \frac{c^3}{16\pi G}\int d^4 x
\sqrt{-{\widetilde
g}}\left\{
{\Phi} {\widetilde R}
-\frac{\omega_{\rm BD}(\Phi)}{\Phi}
\left(\partial_\mu{\Phi}\right)^2
-2 {U(\Phi)}
\right\} + S_{\rm matter}\left[{\rm
matter};
{\widetilde g_{\mu\nu}}\right]\!,~
\label{BDaction}
\end{eqnarray}
in which the scalar curvature $\widetilde R$, the determinant
$\widetilde g$, and the contraction of indices in
$(\partial_\mu{\Phi})^2$ now all correspond to the physical
(``Jordan-frame'') metric $\widetilde g_{\mu\nu}$. Note that this
action defines exactly the \textit{same} theory as
Eq.~(\ref{action}).

The reconstruction equations can then be written straightforwardly.
The first step is to derive the Hubble function $H(z)$ from the
knowledge of the luminosity distance $D_L(z)$:
\begin{equation}
\frac{1}{H(z)} =
\left(\frac{D_L(z)}{1+z}
\right)'\times\left[1+{\Omega_{\kappa,0}}\left(
\frac{H_0 D_L(z)}{1+z}
\right)^2\right]^{-1/2},
\end{equation}
where a prime denotes here a derivation with respect to the redshift
$z$, and $\Omega_{\kappa,0}$ is the possible (small) contribution of
the Universe's spatial curvature to the total energy density. In the
short wavelength limit $\lambda \ll (H^{-1}\,, m_\varphi^{-1})$, the
field equations for small perturbations yield $\ddot\delta_m + 2 H
\dot \delta_m - 4\pi G_{\rm eff} \rho \delta_m\simeq 0$, where
$G_{\rm eff} \equiv G [A^2(\varphi) + (dA/d\varphi)^2]$ is the
effective gravitational constant between two close test masses,
taking into account the scalar-field contribution. In terms of the
Brans-Dicke variables, this gives
\begin{equation}
\frac{\Phi}{\Phi_0} \simeq
{\frac{3}{2}\left(\frac{H_0}{H}\right)^2
\frac{(1+z)\,{\Omega_{m,0}}\,\delta_m}
{\delta_m''+\left(\frac{H'}{H} -
\frac{1}{1+z}\right)\delta_m'}
\times \left(1+\frac{1}{2{\omega_{\rm BD}}+3}\right)}.
\label{Phi}
\end{equation}
On the other hand, the background evolution equations yield
\begin{eqnarray}
\frac{2 U}{(1+z)^2H^2} &=&
{{\Phi''} + \left(\frac{H'}{H}
- \frac{4}{1+z}\right){\Phi'}
+\left[\frac{6}{(1+z)^2} - \frac{2}{1+z}\,\frac{H'}{H}
{- 4
\left(\frac{H_0}{H}\right)^2\!
\Omega_{\kappa,0}}\right]{\Phi}}
\nonumber\\
&& {- 3\,(1+z)\left(\frac{H_0}{H}\right)^2{\Phi_0}\,
{\Omega_{m,0}}}\,,
\label{U}\\
{\omega_{\rm BD}} &=&
{-\frac{\Phi}{\Phi'^2}
\Biggl\{{\Phi''}
+ \left(\frac{H'}{H} + \frac{2}{1+z}\right)
\Phi' -2\left[\frac{1}{1+z}\,\frac{H'}{H}
{- \left(\frac{H_0}{H}\right)^2\!
\Omega_{\kappa,0}}\right]{\Phi}}
\nonumber\\
&& {+ 3\,(1+z) \left(\frac{H_0}{H}\right)^2
{\Phi_0}\,{\Omega_{m,0}}
\Biggr\}}.
\label{omegaBD}
\end{eqnarray}
Therefore, Eqs.~(\ref{Phi})--(\ref{omegaBD}) allow us to reconstruct
${\Phi}(z)$, ${U}(z)$ and ${\omega_{\rm BD}}(z)$ as functions of the
redshift, and thereby ${U}({\Phi})$ and ${\omega_{\rm BD}}({\Phi})$
in a parametric way. Actually, the last factor of Eq.~(\ref{Phi})
may be neglected, since we know that $(2\omega_{\rm BD} +3) ^{-1} =
\alpha_0^2 \ll 1$ at present. In that case,
Eqs.~(\ref{Phi})--(\ref{omegaBD}) become merely algebraic, and no
integration of differential equation is required. However, one
should note that second derivatives of experimental data are needed,
so that large uncertainties are expected in such a reconstruction
program (see R.~Jimenez's contribution to the present proceedings).
The point that we wish to emphasize is that the full reconstruction
of the microscopic action (\ref{BDaction}) is theoretically
possible. Although it needs some algebra, this result seems anyway
obvious: It is possible to \textit{fit} two observed functions
[$D_L(z)$ and $\delta_m(z)$] thanks to two unknown ones
[$V(\varphi)$ and $A(\varphi)$, or equivalently ${U}({\Phi})$ and
${\omega_{\rm BD}}({\Phi})$].

However, future experiments (like the SNAP satellite) will only give
access to the luminosity distance $D_L(z)$ with a good accuracy, and
the density contrast $\delta_m(z)$ cannot yet be used to constrain
the models. A \textit{semi-phenomenological} approach can thus be
useful: We make some theoretical hypotheses on either the potential
$V(\varphi)$ or the coupling function $A(\varphi)$, and we
reconstruct the other one from $D_L(z)$. \textit{A priori}, one
may think that such a reconstruction is again obvious: We fit one
observed function [$D_L(z)$] with one unknown function [$V(\varphi)$
or $A(\varphi)$]. But this naive reasoning is only valid locally, on
a small interval. Indeed, the reconstructed function may for
instance diverge for some value of the redshift, or one of the
degrees of freedom may need to take a negative energy beyond a given
redshift, which would make the theory unstable (and ill defined as a
field theory on the surface where the energy changes its sign). The
positivity of the graviton energy implies $A^2(\varphi) > 0$, which
can be translated in terms of the Brans-Dicke scalar field as $\Phi
> 0$. On the other hand, the positivity of the scalar-field energy
imposes the minus sign in front of the scalar kinetic term
$-(\partial_\mu\varphi)^2$ in action (\ref{action}), which
translates as $\omega_{\rm BD} > -\frac{3}{2}$ in terms of the
Brans-Dicke parameter. It was shown in Ref.~\cite{efp01} that
these conditions impose tight constraints on the theories as soon
as one knows $D_L(z)$ over a wide enough interval $z\in[0,\sim 2]$.

\begin{figure}
\includegraphics[height=.25\textheight]{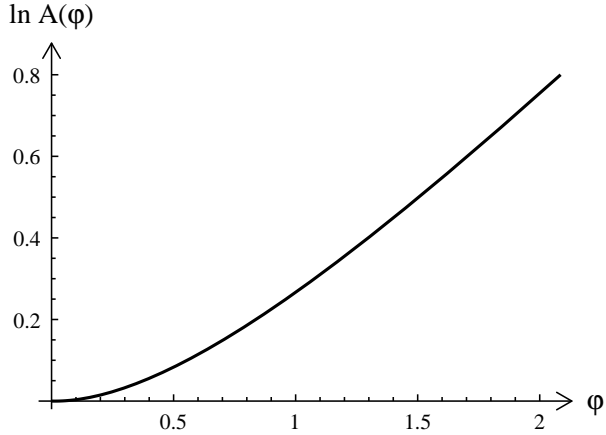}
\caption{Matter-scalar coupling function $\ln A(\varphi)$ which
reproduces exactly the same accelerated expansion as GR plus a
cosmological constant (with $\Omega_{\Lambda} \simeq 0.7$), without
any scalar field potential $V(\varphi)$ nor any cosmological
constant. However, this model works only for redshifts $z \lesssim
0.7$.
\label{fig5}}
\end{figure}
For instance, the present accelerated expansion of the universe can
be perfectly described by a scalar-tensor theory with a vanishing
potential $V(\varphi) = 0$ (and therefore a vanishing cosmological
constant too). Reference~\cite{efp01} derived analytically the
coupling function $A(\varphi)$ which reproduces exactly the same
evolution of the scale factor $a(z)$ as the one predicted by GR plus
a cosmological constant. Figure~\ref{fig5} displays this
reconstructed function $\ln A(\varphi)$, which has a nice parabolic
shape, with a minimum very close to the present value $\varphi_0$ of
the scalar field, and a positive second derivative. This is not only
consistent with binary-pulsar data [which forbid large and negative
values of this second derivative $\beta_0$, see
Eq.~(\ref{beta0min})] but also with the cosmological attractor
phenomenon analyzed by Damour and Nordtvedt~\cite{dn93}: The scalar
field is generically attracted towards a minimum of $\ln A(\varphi)$
during the cosmological expansion, whereas some fine tuning would be
necessary to reach a maximum (negative $\beta_0$). Therefore, we are
in the difficult situation in which two very different theories are
both consistent with experimental data, and there seems to be no way
to distinguish them. Fortunately, Ref.~\cite{efp01} found that this
scalar-tensor theory cannot mimic GR plus a cosmological constant
beyond a redshift $z \sim 0.7$, because the scalar field $\varphi$
would diverge at this value, and above all because the graviton
energy would become negative beyond. Therefore, it suffices to
measure $D_L(z)$ precisely enough up to $z \sim 1$ to rule out such
a potential-free scalar-tensor theory. Actually, if $D_L(z)$ is
measured over a wider interval $z\in[0,\sim 2]$, large experimental
errors (tens of percent) are not problematic: It is still possible
to distinguish this potential-free model from GR plus a cosmological
constant, and thereby to rule out one of them. The results of the
SNAP satellite up to $z \sim 2$ will therefore be very useful to
constrain scalar-tensor theories of gravity.

One can also impose a particular form of the coupling function
$A(\varphi)$ and reconstruct the potential $V(\varphi)$ which
reproduces the observed luminosity distance $D_L(z)$. For instance,
for a minimally coupled scalar field $A(\varphi) = 1$ (usual
``quintessence'') in a spatially curved universe, Ref.~\cite{efp01}
analytically derived the expression of $V(\varphi)$ which gives the
same cosmological evolution as GR plus a cosmological constant in a
spatially flat universe. It was found that the shape of the
potential is smoother when the universe is (marginally) closed. If
it is flat or almost flat, one obviously recovers a cosmological
constant with its unnaturally small value $\Lambda \simeq 3\times
10^{-122} c^3/(\hbar G)$. Therefore, in that case, aesthetic reasons
may help us discriminate between the theories, instead of the much
stronger argument of the positivity of energy that was used above.
This shows anyway that the sole knowledge of $D_L(z)$ suffices to
constrain scalar-tensor theories of gravity.

\section{The most general scalar-tensor theories}
\label{general}
More general scalar-tensor theories than action (\ref{action}) can
be defined. First, one may consider several scalar
fields~\cite{def1}, labeled by latin indices, $\varphi^a$. In the
action, their kinetic term $-\frac{1}{2}\partial_\mu\varphi^{a}
\partial^\mu\varphi^{b}$ will thus involve a matrix $\gamma_{ab}$ to
contract the indices. Such a matrix may itself depend on the scalar
field, and it defines thus what is called a $\sigma$-model. An
example of a tensor-bi-scalar theory is provided by hybrid
inflation. Note that the $\sigma$-model metric $\gamma_{ab}$ must be
positive definite for all the scalar fields to carry positive
energy.

The quintessence potential $V(\varphi^a)$ may also depend on the
various scalar fields, and thereby define $n$-dimensional mountains
along which the scalar degrees of freedom roll down. Similarly, the
matter-scalar coupling function $A(\varphi^a)$ may have a nontrivial
shape, and define a rich ``extended quintessence'' phenomenology.

If the weak equivalence principle is not exactly satisfied,
different species of matter, labeled $i$, may have different
coupling functions $A_i(\varphi^a)$ to the scalar field. It should
be noted that superstring theory does predict such couplings, which
violate equivalence principle tests by 9 or 10 orders of magnitude
at tree level (see J.P.~Uzan's contribution to the present
proceedings). To reconcile it with experiment, the standard argument
is that no symmetry protects the masses of the scalar fields, so
that all of them should acquire a large enough mass at present to be
exponentially suppressed, or anyway a potential such that they are
very weakly coupled to matter on Earth (cf.~the chameleon
model~\cite{Khoury}). However, ``flat directions'' generically arise
in low-energy effective models inspired by superstrings, i.e.,
one or several scalar fields which remain strictly massless and
violate \textit{a priori} the weak equivalence principle. In that
case, string loop corrections can induce an attractor
mechanism~\cite{dp94}, which efficiently drives such scalars towards
an extremum of their coupling functions to matter, so that they
almost decouple at present.

It is also possible to couple the scalar fields to the Gauss-Bonnet
topological invariant, $R_{\mu\nu\rho\sigma}^2 -4 R_{\mu\nu}^2 +
R^2$. This is the only combination involving powers of the Riemann
and/or the Ricci tensors which preserves the spectrum of the theory,
i.e., which does not add any extra degree of freedom. A function
$f(R)$ of the scalar curvature is well known to be equivalent to an
extra scalar degree of freedom in the model, cf.~\cite{teyssandier}.
It suffices to rewrite $\int f(R)$ as $\int \left[f(\varphi) +
(R-\varphi)f'(\varphi)\right]$ and to show that the field equations
deriving from these actions coincide. The latter is of the
Brans-Dicke type (\ref{BDaction}), with $\omega_{\rm BD} = 0$, and a
mere change of variables (conformal transformation) $g^*_{\mu\nu}
\equiv f'(\varphi) g_{\mu\nu}$ yields the canonical
form~(\ref{action}). Similarly, a function $f(R,\Box R, \ldots,
\Box^n R)$ of the scalar curvature and its iterated d'Alembertian
can be shown to be equivalent to the introduction of $n+1$ scalar
fields in the theory~\cite{gottlober}. On the other hand, functions
$f(R_{\mu\nu})$ or $f(R_{\mu\nu\rho\sigma})$ of the Ricci or the
Riemann tensor have been proved several times~\cite{stelle} to
involve generically an extra massive spin-2 ghost, i.e., an extra
negative-energy graviton which spoils the stability of the theory.
The only allowed combination is the Gauss-Bonnet topological
invariant.

Finally, one may also consider a function $F(\partial_\mu\varphi
\partial^\mu\varphi)$ of the kinetic terms, as in ``k-essence''
models~\cite{kessence}. However, only some particular functions
$F$ preserve the positivity of the scalars' kinetic energy.

In conclusion, the most general tensor--multi-scalar theories that
we may consider are defined by the following action:
\begin{eqnarray}
S&=&\frac{c^3}{4\pi G} \int\sqrt{-g} \left\{\frac{R}{4}
-\frac{1}{2}\, F{\Bigl(} g^{\mu\nu} \gamma_{ab}(\varphi^c)\,
\partial_\mu\varphi^{a} \partial_\nu\varphi^{b} \Bigr)
-V(\varphi^{a})\right\}\nonumber\\
&&- \hbar\int\sqrt{-g}\, W(\varphi^{a})
\left(R_{\mu\nu\rho\sigma}^2
-4R_{\mu\nu}^2 + R^2\right)\nonumber\\
&&+\sum_i S_{\mathrm{matter}_i}\left[
\mathrm{matter}_i\,; \widetilde g^{(i)}_{\mu\nu}
\equiv A_i^2(\varphi^{a})g_{\mu\nu} \right].
\end{eqnarray}

\section{Experimental constraints on a scalar--Gauss-Bonnet
coupling}
\label{GaussBonnet}
In this section, we will consider the very particular case of a
scalar-tensor theory with a single scalar field $\varphi$,
no matter-scalar coupling [$A(\varphi) = 1$], no potential
[$V(\varphi) = 0$], and a standard kinetic term $-\frac{1}{2}
(\partial_\mu\varphi)^2$, but with a scalar--Gauss-Bonnet
coupling $W(\varphi)$:
\begin{eqnarray}
S&=&\frac{c^3}{4\pi G}
\int\sqrt{-g}
\left\{\frac{R}{4}
-\frac{1}{2}\left(\partial_\mu{\varphi}\right)^2\right\}
- \hbar\int\sqrt{-g}\,
{W(\varphi)}
\left({R_{\mu\nu\rho\sigma}^2
-4R_{\mu\nu}^2 + R^2}\right)\nonumber\\
&&+S_\mathrm{matter}\left[
\mathrm{matter}\,; g_{\mu\nu} \right].
\label{GBaction}
\end{eqnarray}
Our aim is just to illustrate the kind of constraints which can be
imposed on such a scalar--Gauss-Bonnet coupling~\cite{moriond03}.
This model will underline that solar-system and cosmological
observations must be sometimes simultaneously taken into account
to constrain a theory. The perturbative but highly nonlinear effects
which happen can also be instructive.

Solar-system and binary-pulsar tests are local, and any deviation
{}from GR depends on the magnitude of the scalar field created by a
massive body. Let us thus analyze first the equation satisfied by
$\varphi$ in the vicinity of a spherical mass $m_\odot$. We can
assume that the metric is close to the Schwarzschild solution, and
we get at the first nonvanishing order in powers of $Gm_\odot/c^2$
\begin{equation}
\Box \varphi = \frac{3 \ell_0^2}{r^6}
\left(\frac{2 G m_\odot}{c^2}\right)^2
\left[W'_0 + W''_0 \varphi + \mathcal{O}(\varphi^2)\right]\ ,
\label{boxphi}
\end{equation}
where we have set $\ell_0^2 \equiv 16 \pi G \hbar / c^3$, and where
the derivative $W'(\varphi)$ has been expanded in powers of
$\varphi$ in the right-hand side. Since we are assuming that
$\varphi$ takes small values, let us neglect the contribution $W''_0
\varphi$. We can then compute any observable prediction, but we
quote below only the results for the light deflection angle
($\Delta\theta_*$) and for the perihelion shift per orbit
($\Delta\theta_p$), which suffice for our purpose:
\begin{eqnarray}
\Delta\theta_* &=& \frac{4 G m_\odot}{r_0 c^2} +
\frac{1536}{35}\left(\frac{G m_\odot}{r_0 c^2}\right)^3
\left(\frac{\ell_0}{r_0}\right)^4 W'^2_0\ ,
\label{defl0}\\
\Delta\theta_p &=& \frac{6\pi G m_\odot}{p c^2} +
192\pi\left(\frac{G m_\odot}{p c^2}\right)^2
\left(\frac{\ell_0}{p}\right)^4 W'^2_0\ ,
\label{peri0}
\end{eqnarray}
where $r_0$ is the minimal distance between the light ray and the
Sun, and $p$ is the \textit{semilatus rectum} of an orbit. The first
terms on the right-hand sides are the usual GR predictions,
at first order in $G m_\odot/c^2$. In conclusion,
solar-system (and binary-pulsar) tests can easily be passed if
$|W'_0|$ is small enough.

One can now reconstruct the full shape of $W(\varphi)$ from the
cosmological observation of the luminosity distance $D_L(z)$, as in
Sec.~\ref{reconstruction} above. We find that this can always
been done, without any problem of negative energy, contrary to
what happened above for the matter-scalar coupling function
$A(\varphi)$. Moreover, there exists again an attraction
mechanism which drives the scalar field towards a minimum of
$W(\varphi)$ during the cosmological expansion. Therefore, a small
value of the slope $|W'_0|$ is indeed expected at present,
consistently with what is needed for solar-system tests. In
conclusion, we are faced again with a serious problem: We just found
a theory which seems to be consistent with all experimental data,
although it is very different from GR in its field content. Our aim
is therefore to find a way to distinguish it from GR, or to rule it
out for internal consistency reasons.

\begin{figure}
\includegraphics[height=.3\textheight]{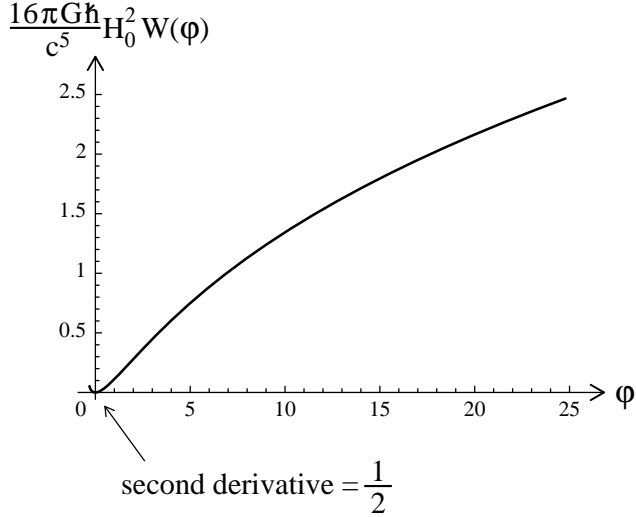}
\caption{Scalar--Gauss-Bonnet coupling function $W(\varphi)$ which
exactly reproduces the cosmological expansion predicted by GR plus a
cosmological constant.
\label{fig6}}
\end{figure}
Figure \ref{fig6} displays this reconstructed coupling function
$W(\varphi)$, in which the present value of the scalar field is
close to the minimum at $\varphi=0$. Its shape is nicely smooth, but
its second derivative at the origin is huge if one divides it by the
tiny factor $(16\pi G\hbar/c^5) H_0^2$. One gets $W''_0 \simeq
7\times 10^{119}$, which is in fact not surprizing, since the
coupling function $W(\varphi)$ behaves in action (\ref{GBaction}) as
the inverse of a cosmological constant. Indeed, $W(\varphi)$
multiplies the square of the curvature tensor, whereas the usual
Einstein-Hilbert term involves the first power of the scalar
curvature, and a cosmological constant does not multiply any
curvature term at all. It was thus expected that $W(\varphi)$
involve a dimensionless number of the order of the inverse of
$(\hbar G/c^3)\Lambda \simeq 3\times 10^{-122}$. Therefore, this
model is ugly, but it is not yet ruled out. One should not confuse
fine tuning and large (or small) dimensionless numbers in a model.
We are here in the second situation, but there is \textit{a priori}
no fine tuning since the scalar field is attracted towards the
minimum of $W(\varphi)$ during the cosmological expansion. There
remains to study how efficiently it is attracted, but this is
actually not necessary for our purpose.

Indeed, $W''_0$ takes such a gigantic value that an approximation
that we made to analyze solar-system tests is no longer valid.
Indeed, we have $|W''_0\varphi| \gg |W'_0|$, so that the second
term on the right-hand side of Eq.~(\ref{boxphi}) cannot be
neglected. To simplify the discussion, we will anyway assume that
$W(\varphi)$ is parabolic, which is a good approximation in a
vicinity of the minimum $\varphi = 0$. We will thus neglect the
higher order terms $\mathcal{O}(\varphi^2)$ in Eq.~(\ref{boxphi});
taking them into account would not change our conclusions below. We
did not find a close analytic solution to Eq.~(\ref{boxphi}), but is
is possible to write it as a series
\begin{eqnarray}
\varphi&=&\frac{W'_0}{W''_0}
\sum_{n\geq 1}\frac{1}{(3\times 4)(7\times 8)\cdots (4n-1)(4n)}
\left(\frac{12 \ell_0^2 G^2 m_\odot^2 W''_0}{r^4 c^4}\right)^n
\label{phiexact}\\
&\simeq&\frac{W'_0}{W''_0}
\left[
\vcenter{\hbox{$\cos$}\hbox{$\cosh$}}
\left(\frac{G m_\odot \ell_0}{r^2 c^2}
\sqrt{3|W''_0|}\right)-1\right]\qquad
\vcenter{\hbox{if $W''_0<0$,}\hbox{if $W''_0>0$.}}
\label{phiapprox}
\end{eqnarray}
The second expression is a good approximation if the argument of the
cosine (or hyperbolic cosine if $W''_0 > 0$) is much greater than 1.
This is the case if we use the huge value of $W''_0$ obtained above
{}from the cosmological reconstruction, and a typical solar-system
distance for the radius $r$: The argument of the hyperbolic cosine
is then of order $10^8$.

The above solution is such that $\varphi \propto W'_0$, therefore we
do not find any nonperturbative effect similar to the ``spontaneous
scalarization'' of neutron stars mentioned in Sec.~\ref{nonpert}
above. Moreover, $\varphi \rightarrow 0$ as $ r \rightarrow \infty$,
and we recover GR for distances $r > 4\times 10^{14}~\rm{m}$ (i.e.,
farther that the solar system including Oort's comet cloud). On the
other hand, there are highly nonlinear corrections proportional to
$1/r^{4n}$ within the solar system. Since the ratio $(12 \ell_0^2
G^2 m_\odot^2 W''_0/ r^4 c^4)$ is much greater that $1$, its
successive powers blow up, but they are compensated by the factors
$1/(3\times4\times7\times\cdots 4n)$ which behave like the inverse
of factorials. Therefore, the successive terms of series
(\ref{phiexact}) start to grow exponentially, then reach a maximum
for a value of the index $n$ which may be large, and finally tend
towards zero. Each of these successive terms must be assumed to be
small enough for the model to pass all classical tests, but one
should not forget that the largest one does not correspond to $n=1$.

In order to study the effects of such highly nonlinear terms in the
solar system, we compute their corrections to the Schwarzschild
metric in the form
\begin{equation}
ds^2 = -\left(1+\sum_{n\geq 1}\frac{\beta_n}{r^n}\right)c^2 dt^2 +
\left(1+\sum_{n\geq 1}\frac{\alpha_n}{r^n}\right)dr^2 +r^2
\left(d\theta^2 +\sin^2\theta\, d\phi^2\right)\ ,
\end{equation}
and we find that the light deflection angle and the perihelion shift
are respectively given by
\begin{eqnarray}
\Delta\theta_* &=& \sum_{n\geq 1} 2^{n-1}\frac{\Gamma\left(
\frac{n+1}{2}\right)^2}
{\Gamma(n+1)}\,\frac{\alpha_n-n\beta_n}{r_0^n} +
\mathcal{O}(\alpha_n,\beta_n)^2\ ,
\label{defl}\\
\Delta\theta_p &=& \frac{6\pi G m_\odot}{p c^2} -
\sum_{n\geq 3} \frac{n(n-1)\beta_n c^2}{2 Gm_\odot p^{n-1}}\,\pi +
\mathcal{O}(\alpha_n,\beta_n)^2\ ,
\label{peri}
\end{eqnarray}
which generalize Eqs.~(\ref{defl0})-(\ref{peri0}) above. Note that
these results are (independently) perturbative in each of the
coefficients $\alpha_n$ and $\beta_n$, because we are assuming that
the scalar-field effects are negligible with respect to the general
relativistic predictions. However, the dominant scalar-field
corrections may correspond to a large value of index $n$.

When solution (\ref{phiexact}) or its approximation
(\ref{phiapprox}) are used to compute the metric coefficients
$\alpha_n$ and $\beta_n$ in the above observable predictions,
and if we use the huge value of $W''_0$ obtained from the previous
cosmological reconstruction of $W(\varphi)$, we get the following
experimental constraint:
\begin{equation}
|W'_0| < 10^{-2\times 10^{11}}.
\end{equation}
Now we can speak of fine tuning, and even of \textit{hyperfine}
tuning! This constraint simply means that the present value of the
scalar field must be \textit{exactly} at the minimum of the coupling
function $W(\varphi)$, otherwise solar-system tests are violated.
And since the universe is still evolving, the scalar field cannot
remain so close to the minimum for more than a fraction of a second.
Therefore, even if we assumed that $W'_0 = 0$ strictly to pass
solar-system tests, this would not be the case a tiny instant later.
In conclusion, we managed to rule out the scalar-tensor model
$A(\varphi) = 1$, $V(\varphi) = 0$ and $W(\varphi) \neq 0$. It
cannot describe an accelerating expansion of the universe at present
and pass solar-system (and binary-pulsar) tests at the same time.

Of course, this result does not rule out any scalar--Gauss-Bonnet
coupling. A model with three (or even two) free functions
$A(\varphi)$, $V(\varphi)$ and $W(\varphi)$ can obviously pass all
present tests. [For instance, GR plus a cosmological constant simply
corresponds to $A(\varphi)=1$, $V(\varphi)=\Lambda/2$ and
$W(\varphi)=0$.] But the presence of a non-constant coupling
$W(\varphi)$ can change the physics at small scales, notably in the
very early universe (Big-Bang) and for later clustering properties.

The fact that $W(\varphi)$ induces effects at small scales can be
understood by a simple dimensional argument. Since this function
multiplies the square of the curvature in action (\ref{GBaction}),
it induces corrections proportional to $1/r^7$ (and higher orders)
to the Newtonian potential in $1/r$, and thereby generically
dominates at small scales. However, we saw above that this quick
reasoning can be erroneous in some perturbative but highly nonlinear
situations. Indeed, if $W''_0$ takes very large and
\textit{negative} values, the cosine involved in
Eq.~(\ref{phiapprox}) shows that $\varphi$ is always of the order of
$-W'_0/W''_0$, even for small distances $r$. One can then prove that
the (very easily satisfied) condition $|\ell_0^2 W'_0| \ll r^2$
suffices for all scalar-field effects to be negligible in the solar
system, even if $|W''_0| \sim 10^{120}$. This remark underlines that
nonlinear effects can drastically change the intuitive behavior, but
let us recall that our cosmological reconstruction above predicted a
large and \textit{positive} value for $W''_0$. In that case, we did
find that the scalar--Gauss-Bonnet coupling induces large effects at
small scales, and even exponentially larger than the linear results
(\ref{defl0})-(\ref{peri0}).

\section{Conclusions}
\label{conclusions}
Scalar-tensor theories of gravity are the best motivated
alternatives to general relativity. Three classes of experimental
data give \textit{qualitatively} different constraints on them.
Solar-system tests strongly constrain the first derivative of the
matter-scalar coupling function $A(\varphi)$ (i.e., the linear
matter-scalar coupling strength $\alpha_0$). Binary-pulsar data
forbid large and negative values of its second derivative $\beta_0$
(quadratic matter-scalar-scalar coupling). The knowledge of the two
cosmological functions $D_L(z)$ and $\delta_m(z)$ suffices to
reconstruct the full shape of both $A(\varphi)$ and the potential
$V(\varphi)$ on a finite interval of $\varphi$. The knowledge of the
luminosity distance $D_L(z)$ alone over a \textit{wide} redshift
interval strongly constrains the theories if one takes into account
solar-system (and binary-pulsar) data, the positivity of the
graviton and scalar energies, and the stability and naturalness of
the models. Future data, provided by experiments like the SNAP
satellite, will notably allow us to discriminate between GR plus a
cosmological constant and a potential-free scalar-tensor theory. The
possible coupling $W(\varphi)$ of the scalar field to the
Gauss-Bonnet topological invariant can be constrained only if one
takes into account cosmological and solar-system data together. The
predictions of the model at small distances can depend on highly
nonlinear corrections. Of course, a model including all three
functions $A(\varphi)$, $V(\varphi)$ and $W(\varphi)$ is
experimentally allowed, since GR plus a cosmological constant is a
particular case. The presence of a scalar--Gauss-Bonnet coupling
$W(\varphi)$ will generically change the behavior of the theory at
small scales (clustering, Big Bang). More general
tensor--multi-scalar theories, including couplings of the scalars to
matter and the Gauss-Bonnet invariant, and including non-quadratic
``k-essence'' kinetic terms, can lead to a much richer
phenomenology, but at the price of a less natural Lagrangian.

\acknowledgments
I wish to thank the organizers of the workshop for
their kind invitation, and my colleagues T.~Damour, D.~Polarski,
E.~Semboloni, A.A.~Starobinsky, and J.P.~Uzan for collaborations
or discussions on this subject.

\end{document}